\documentclass[twocolumn,floats,prb,aps,showpacs]{revtex4}
\usepackage{graphicx}
\usepackage{amsfonts}
\usepackage{amsmath}
\usepackage{amssymb}
\usepackage{exscale}
\usepackage{color}

\begin{document}

\title{Two-Cooper-pair problem and the Pauli exclusion principle}
\author{Walter V. Pogosov$^{1,2}$, Monique Combescot$^{1}$, Michel Crouzeix$%
^{3}$}
\affiliation{(1) Institut des NanoSciences de Paris, Universit\'e Pierre et Marie Curie,
CNRS, Campus Boucicaut, 140 rue de Lourmel, 75015 Paris}
\affiliation{(2) Institute for Theoretical and Applied Electrodynamics, Russian Academy
of Sciences, Izhorskaya 13, 125412 Moscow}
\affiliation{(3) Institut de Recherche Math\'ematiques de Rennes, Universit\'e de Rennes
1, Campus de Beaulieu, 35042 Rennes cedex, France}
\date{\today }

\begin{abstract}
While the one-Cooper pair problem is now a textbook exercise, the energy of
two pairs of electrons with opposite spins and zero total momentum has not
been derived yet, the exact handling of Pauli blocking between bound pairs
being not that easy for $N=2$ already. The two-Cooper pair problem however
is quite enlightening to understand the very peculiar role played by the
Pauli exclusion principle in superconductivity. Pauli blocking is known to
drive the change from 1 to $N$ pairs, but no precise description of this
continuous change has been given so far. Using Richardson's procedure, we
here prove that Pauli blocking increases the free part of the two-pair
ground state energy, but decreases the binding part when compared to two
isolated pairs - the excitation gap to break a pair however increasing from
one to two pairs. When extrapolated to the dense BCS regime, the decrease of
the pair binding while the gap increases strongly indicates that, at odd
with common belief, the average pair binding energy cannot be of the order
of the gap.
\end{abstract}

\author{}
\maketitle
\date{\today }

\section{Introduction}

The first step towards understanding the microscopic grounds of
superconductivity was made by Fr\"{o}hlich\cite{Frol} who has realized that
electrons in metals can form bound pairs due to their weak interaction with
the ion lattice, which results in an effective electron-electron attraction.
A few years later, Cooper has considered\cite{Cooper} a simplified quantum
mechanical problem of two electrons with opposite spins and zero total
momentum added to a "frozen" Fermi sea, i.e., a sea of noninteracting
electrons. Within the Cooper model, an attractive interaction between these
two electrons is introduced, this interaction being localized in a
finite-width layer above the "frozen" Fermi sea. Cooper has shown that such
an attraction, no matter how weak, leads to the appearance of a bound state
for the two additional electrons. This result was demonstrated for a single
pair although it was fully clear that conventional superconductivity takes
place in a macroscopic system of electrons paired by such an attraction.

One year later, Bardeen, Cooper and Schrieffer\cite{BCS} (BCS) have proposed
an approximate solution of the quantum many-body problem for electrons with
opposite spins attracting each other. A very important result of the BCS
theory is the existence of a gap in the excitation spectrum above the ground
state. In the BCS model, the potential layer, in which an attraction between
electrons with opposite spins acts, extends symmetrically on both sides of
the Fermi level. This implies that indeed a macroscopic number of electrons
interact with each other. In order to avoid the difficult problem associated
with the Pauli exclusion principle between a given number of same spin
electrons, the grand canonical ensemble was used. The original formulation
of BCS theory is also based on a variational ansatz for the ground state
wave function: The wave function is taken with all the electrons feeling the
attraction, paired, i.e., "condensed" into the \textit{same}
quantum-mechanical state.

It was, however, emphasized by Schrieffer that electron pairs are not
elementary bosons because they are constructed from two elementary fermions%
\cite{Schrieffer}, so that their creation and destruction operators do not
obey simple bosonic commutation relations. Schrieffer also claimed that the
large overlap which exists between pairs in the dense BCS configuration cuts
any link with the two-body Cooper model, the isolated pair picture thus
having little meaning in the dense regime\cite{Schrieffer}. In spite of this
claim, it is rather obvious that the many-electron BCS configuration can be
reached from the one-Cooper pair limit by simply adding more and more
electron pairs into the layer where the attraction acts, until the layer
becomes half-filled. A canonical procedure of this kind would allow one to
see the evolution of correlated electron pairs from the dilute to the dense
regime and to understand deeper the role of the Pauli exclusion principle in
fermion pairing. Notice that such an approach can also be considered as a
useful and well-defined toy model for the crossover between local and
extended pairs of attracting fermions, which in the present time attracts
large attention within the field of ultracold gases\cite%
{roland,review1,review2}. The crossover problem is still open even for the
simplest case of the "reduced" BCS potential for fermion-fermion
interaction: a variational solution has only been proposed long time ago by
Eagles\cite{Eagles} and also by Leggett\cite{Tonycross}. It also uses a
BCS-like ansatz for the ground state wave function.

A possible way to tackle the problem in the canonical ensemble, i.e., for a
fixed number of electron pairs, is to use the procedure developed by
Richardson \cite{Rich1, Rich2}. It allows us to write the form of the 
\textit{exact} $N$-pair eigenstate of the Schr\"{o}dinger equation in the
case of the so-called "reduced" BCS potential which is the simplest
formulation of the electron-electron interaction mediated by the ion motion.
The eigenstate, as well the energy of $N$ pairs, read in terms of $N$
parameters $R_{1}$, ..., $R_{N}$, which are solutions of $N$ nonlinear
algebraic equations. Although Richardson's approach greatly simplifies the
problem by avoiding a resolution of a $N$-body Schr\"{o}dinger equation, the
solution of these equations for $R_{1}$, ..., $R_{N}$ in a compact form for
general $N$ remains an open problem. One of the difficulties is due to the
fact that $N$ is not a parameter in these equations but only enters through
the number of equations. This is rather unusual and makes the $N$-dependence
of the system energy quite uneasy to extract. Nowadays, Richardson's
equations are tackled numerically for small-size superconducting granules
containing countable numbers of electron pairs\cite{review}. We wish to add
that the canonical approach has also been used in the form of a variational
fixed-$N$ projected BCS-like theories, see e.g. Ref \cite{delft}.

The goal of the present paper, is to extend the original Cooper's work for 
\textit{one} electron pair to \textit{two} pairs: we \textit{analytically}
solve the two Richardson's equations in the large sample limit. Our work can
be considered as an initial step towards the establishing of the precise
link which exists between dilute and dense regimes of pairs, since it
indicates a general trend for the evolution of the ground state energy with
the increase of pair number, i.e., overlap between pairs. Richardson's
equations are here solved by three methods. They of course give identical
results but shine different light on these equations. The approaches to
tackle Richardson's equations, proposed in this paper, in fact constitute
perspectives for the extension to a larger number of pairs and hopefully to
the thermodynamical limit.

The solution we obtain shows that the average pair binding energy is \textit{%
smaller} in the two-pair configuration than for one pair. This result can be
physically understood by noting that electrons which are paired are
fermions; therefore, by increasing the number of pairs, we decrease the
number of states in the potential layer available to form these paired
states. The energy decrease we here find is actually quite general for
composite bosons\cite{Monique}.

However, extrapolation of this understanding to the dense BCS configuration
faces difficulty within the common understanding of BCS results. Indeed, it
is generally believed\cite{LP,Fetter} that the pair binding energy in the
dense BCS limit is of the order of the superconducting gap $\Delta $. At the
same time, this gap is found as exponentially larger than the single pair
binding energy obtained by Cooper. According to the tendency we here
revealed, the average pair binding energy in the dense regime should be
smaller than that in the one-pair problem.

This discrepancy motivated us to focus on what is called pair binding energy
and more generally "Cooper pair" in the various understanding of the BCS
theory. Usually, pairs are said to have a binding energy of the order of $%
\Delta $. However, such pairs are introduced not ab initio, but to provide a
physical understanding of the BCS result for the ground state energy\cite%
{Fetter,Tinkham}. Pairs with energy of the order of the gap are called
"virtual pairs" by Schrieffer\cite{Schrieffer}. They represent couples of
electrons \textit{excited} above the normal Fermi level for noninteracting
electrons, as a result of the attraction between up and down spin electrons.
Since the Fermi level is smeared out on a scale of $\Delta $ by the
attraction, the number of such pairs is much smaller than the total number
of electron pairs feeling the attraction. The latter were named "superfluid
pairs" by Schrieffer\cite{Schrieffer}. By construction, the concept of
"virtual pair" breaks a possible continuity between the dilute and dense
regimes of pairs in a somewhat artificial way. By contrast, staying within
the framework of "superfluid pairs" greatly facilitates the physical
understanding of the role of Pauli blocking in superconductivity as well as
in the BEC-BCS crossover problem. Our results in fact demonstrate the
importance of a clear separation between the various concepts of "Cooper
pair" found in the literature.

We wish to mention that the results presented in this paper do not have
straightforward experimental applications. The main goal of this paper is to
reveal the general trend for the evolution of energy spectrum when changing
the number of pairs and to make a first step towards a fully controllable
resolution of the $N$-pair problem. However, even a two-pair configuration
has a relation to real materials having correlated pairs of fermions,
because this configuration corresponds to a dilute regime of pairs, realized
in some systems. Conceptually, the overlap between pairs can be tuned either
by changing fermion-fermion interaction or total number of pairs.\textbf{\ }%
We here show that by increasing the overlap between pairs, we block more and
more states available for the construction of paired states. For the first
time, a dilute regime of pairs was addressed by Eagles \cite{Eagles} in the
context of superconducting semiconductors having a low carrier
concentration. In particular, it was shown in this paper that the excitation
spectrum in the dilute regime is controlled by the binding energy of an
isolated pair (in agreement with our results) rather than by a more
cooperative gap which appears, when pairs start to overlap. Thus, this
picture is quite similar to the isolated-pair model considered by Cooper.

There also is a variety of unconventional superconductors which are
characterized by rather short coherence length that implies pairs not
overlapping so strongly as in conventional low-$T_{c}$ materials. For
instance, it was argued in Ref. \cite{Randeria} that the BEC-BCS crossover
might be relevant for high-$T_{c}$ cuprates. Some experiments seem to
support this idea, for example Ref \cite{Uemura} where experimental data on
the dependence of the superconducting transition temperature on Fermi
temperature are collected for various superconducting materials. This
analysis indicates that conventional low-$T_{c}$ superconductors stay apart
from short-coherence length materials, including heavy fermion
superconductors. Thus, it was argued that to understand these unconventional
materials, it is appropriate to focus on the most basic aspect, i.e., on the
short coherence length, rather than to introduce more exotic and less
generic concepts \cite{Levin}. It was also shown that the very recently
discovered Fe-based pnictides, which constitute a new class of high-$T_{c}$
superconductors, should be understood as low-carrier density metals
resembling underdoped cuprates\cite{Rosner}, so that it is possible that the
BEC-BCS crossover phenomenon is relevant for these materials as well. Quite
recently, it was demonstrated in Ref. \cite{Shanenko} that size quantization
in nanowires made of conventional superconductors can result in a dramatic
reduction of the coherence length bringing superconducting state to the
BEC-BCS crossover regime. We finally would like to mention that the
two-correlated pair problem has received great attention within the
ultracold gas field, see e.g. Ref. \cite{roland}. All these examples
demonstrate that, paradoxically, the Cooper problem seems to be more
relevant to modern physics than several decades ago. It is also worth
mentioning that BCS Hamiltonian, which only includes interaction between the
up and down spin electrons with zero pair momentum, is oversimplified.
Nevertheless, fermionic pairs in the BEC-BCS transition regime have not been
described yet in a fully controlled manner even within this Hamiltonian. 
\textit{One}\ of the possible strategies to tackle this crossover therefore
is to find a precise solution of the problem for the simplest Hamiltonian
and only after that, to turn to more elaborate Hamiltonians.

The paper is organized as follows. In Section II, we briefly recall the
one-Cooper pair problem to settle notations. In Section III, we present two
solutions to the two-pair ground state, as well as a discussion of the
possible excited states. We conclude in Section IV. In the Appendix, we give
another exact solution to the two-pairs Richardson's equations which shines
a different light to the problem.

\section{The one-Cooper pair problem}

Let us briefly recall the one-Cooper pair problem. We consider a Fermi sea $%
\left\vert F_{0}\right\rangle $ made of electrons with up and down spins. An
attractive potential between electrons with opposite spins and opposite
momenta acts above the Fermi level $\varepsilon _{F_{0}}$. This potential is
taken as constant and separable to allow analytical calculations. In terms
of free pair creation operators $\beta _{\mathbf{k}}^{\dagger }=a_{\mathbf{k}%
\uparrow }^{\dagger }a_{-\mathbf{k}\downarrow }^{\dagger }$\ it reads as%
\begin{equation}
\mathcal{V}=-V\sum_{\mathbf{k}^{\prime },\mathbf{k}}w _{\mathbf{k}^{\prime
}}w _{\mathbf{k}}\beta _{\mathbf{k}^{\prime }}^{\dagger }\beta _{\mathbf{k}}
\label{1}
\end{equation}%
$V$\ is a positive constant and $w _{\mathbf{k}}=1$ for $\varepsilon
_{F_{0}}<\varepsilon _{k}<\varepsilon _{F_{0}}+\Omega $.

We add a pair of electrons with opposite spins to the "frozen" sea $%
\left\vert F_{0}\right\rangle $. When the pair has a nonzero momentum, it is
trivial to see that $a_{\mathbf{p}\uparrow }^{\dagger }a_{-\mathbf{p^{\prime
}}\downarrow }^{\dagger }\left\vert F_{0}\right\rangle $ with $\mathbf{p}%
\neq \mathbf{p^{\prime }}$ is eigenstate of $H=H_{0}+\mathcal{V}$ where $%
H_{0}=\sum_{\mathbf{k,}s}\varepsilon _{k}a_{\mathbf{k}s}^{\dagger }a_{%
\mathbf{k}s}$, its energy being $(\varepsilon _{p}+\varepsilon _{p^{\prime
}})$. If the pair has a zero total momentum, the $H$ eigenstates are linear
combinations of $\beta _{\mathbf{k}}\left\vert F_{0}\right\rangle $. We look
for them as 
\begin{equation}
\left\vert \psi _{1}\right\rangle =\sum_{\mathbf{k}}G(\mathbf{k})\beta _{%
\mathbf{k}}^{\dagger }\left\vert F_{0}\right\rangle  \label{2}
\end{equation}%
%
%
%
%
%
%
%
%
%
%
%
%
%
%
%
%
%
%
%
%
%
%
The Schr\"{o}dinger equation $\left( H-\mathcal{E}_{1}\right) \left\vert
\psi _{1}\right\rangle =0$ imposes $G(\mathbf{k})$ to be such that 
\begin{equation}
\left[ 2\varepsilon _{\mathbf{k}}-\mathcal{E}_{1}\right] G(\mathbf{k})-Vw_{%
\mathbf{k}}\sum w_{\mathbf{k}^{\prime }}G(\mathbf{k}^{\prime })=0  \label{3}
\end{equation}%
For $2\varepsilon _{\mathbf{k}}\neq \mathcal{E}_{1}$, the eigenfunction $G(%
\mathbf{k})$ depends on $\mathbf{k}$ as $w_{\mathbf{k}}/(2\varepsilon _{%
\mathbf{k}}-\mathcal{E}_{1})$, so that $\left\vert \psi _{1}\right\rangle $
is only made of pairs within the potential layer, as physically expected.
The eigenvalues such that $2\varepsilon _{\mathbf{k}}\neq \mathcal{E}_{1}$
for all $\mathbf{k}$ within the potential layer then follows from Eq.(3) as 
\begin{equation}
1=V\sum_{\mathbf{k}}\frac{w_{\mathbf{k}}}{2\varepsilon _{\mathbf{k}}-%
\mathcal{E}_{1}}\simeq \frac{V\rho _{0}}{2}\int_{\varepsilon
_{F_{0}}}^{\varepsilon _{F_{0}}+\Omega }\frac{2d\varepsilon }{2\varepsilon -%
\mathcal{E}_{1}}  \label{4}
\end{equation}%
$\rho _{0}$\ is the mean density of states in the potential layer. This
leads for a weak potential, i.e., a dimensionless parameter $v=\rho _{0}V$
small compared to 1, to 
\begin{equation}
\mathcal{E}_{1}\simeq 2\varepsilon _{F_{0}}-\varepsilon _{c}  \label{5}
\end{equation}%
\begin{equation}
\varepsilon _{c}\simeq 2\Omega e^{-2/v}  \label{6}
\end{equation}%
%
%
%
%
%
%
%
%
%
%
%
%
%
%
%
%
%
%
%
%
As seen below, it will be physically enlightening to rewrite this one-Cooper
pair binding energy as 
\begin{equation}
\varepsilon _{c}\simeq (\rho _{0}\Omega )(2e^{-2/v}/\rho _{0})=N_{\Omega
}\varepsilon _{V}  \label{7}
\end{equation}%
$N_{\Omega }=\rho _{0}\Omega $ is the number of empty pair states in the
potential layer $\Omega $ from which the Cooper pair bound state is
constructed, these states being all empty in the one-Cooper pair problem. $%
\varepsilon _{V}=2e^{-2/v}/\rho _{0}$ appears as a binding energy unit
induced by \textit{each} of the empty pair states in the potential layer. $%
\varepsilon _{V}$ only depends on the potential amplitude $V$ and the
density of states $\rho _{0}$ in the potential layer.

Eq.(7) already shows that the wider the potential layer $\Omega $, the
larger the number of empty states feeling the potential from which the
Cooper pair is made and, ultimately, the larger the binding energy $%
\varepsilon _{c}$. We can also note that the pair binding energy depends 
\textit{linearly} on the number of states available to form a bound state.
This remark is actually crucial to grasp the key role played by Pauli
blocking in superconductivity: Indeed, this blocking makes the number of
empty states available to form a bound state decrease when the pair number
increases - or when one pair is broken as in the case of excited states.

\section{The two-Cooper pair problem}

We now add two pairs having opposite spin electrons and zero total momentum
to the Fermi sea $\left\vert F_{0}\right\rangle $ and we look for the
eigenstates $\left( H-\mathcal{E}_{2}\right) \left\vert \psi
_{2}\right\rangle =0$ as%
\begin{equation}
\left\vert \psi _{2}\right\rangle =\sum G(\mathbf{k}_{1},\mathbf{k}%
_{2})\beta _{\mathbf{k}_{1}}^{\dagger }\beta _{\mathbf{k}_{2}}^{\dagger
}\left\vert F_{0}\right\rangle  \label{8}
\end{equation}%
The bosonic character of fermion pairs which leads to $\beta _{\mathbf{k}%
_{1}}^{\dagger }\beta _{\mathbf{k}_{2}}^{\dagger }=\beta _{\mathbf{k}%
_{2}}^{\dagger }\beta _{\mathbf{k}_{1}}^{\dagger }$, allows us to enforce $G(%
\mathbf{k}_{1},\mathbf{k}_{2})=G(\mathbf{k}_{2},\mathbf{k}_{1})$ without any
lost of generality. The Schr\"{o}dinger equation fulfilled by $G(\mathbf{k}%
_{1},\mathbf{k}_{2})$ is somewhat more complicated than for one pair. To get
it, it is convenient to note that 
\begin{eqnarray}
\mathcal{V}\beta _{\mathbf{k}_{1}}^{\dagger }\beta _{\mathbf{k}%
_{2}}^{\dagger }\left\vert F_{0}\right\rangle &=&-V(1-\delta _{\mathbf{k}_{1}%
\mathbf{k}_{2}})\left( w_{\mathbf{k}_{1}}\beta _{\mathbf{k}_{2}}^{\dagger
}+w_{\mathbf{k}_{2}}\beta _{\mathbf{k}_{1}}^{\dagger }\right)  \notag \\
&&\sum w_{\mathbf{p}}\beta _{\mathbf{p}}^{\dagger }\left\vert
F_{0}\right\rangle  \label{9}
\end{eqnarray}%
the factor $(1-\delta _{\mathbf{k}_{1}\mathbf{k}_{2}})$ being necessary for
both sides of the above equation to cancel for $\mathbf{k}_{1}=\mathbf{k}%
_{2} $. When used into $\left( H-\mathcal{E}_{2}\right) \left\vert \psi
_{2}\right\rangle =0$ projected upon $\left\langle F_{0}\right\vert \beta _{%
\mathbf{k_{1}}}\beta _{\mathbf{k_{2}}}$ we get%
\begin{eqnarray}
&&0=(1-\delta _{\mathbf{k}_{1}\mathbf{k}_{2}})\left[ \left( 2\varepsilon _{%
\mathbf{k}_{1}}+2\varepsilon _{\mathbf{k}_{2}-}\mathcal{E}_{2}\right) G(%
\mathbf{k}_{1},\mathbf{k}_{2})\right.  \notag \\
&&\left. -V\left( w_{\mathbf{k}_{1}}\sum_{\mathbf{k}\neq \mathbf{k}_{2}}w_{%
\mathbf{k}}G(\mathbf{k},\mathbf{k}_{2})+(\mathbf{k}_{1}\leftrightarrow 
\mathbf{k}_{2})\right) \right]  \label{10}
\end{eqnarray}%
The above equation makes $G(\mathbf{k}_{1},\mathbf{k}_{1})$ undefined. This
however is unimportant since the $\mathbf{k}_{1}=\mathbf{k}_{2}$
contribution to $\left\vert \psi _{2}\right\rangle $ anyway cancels due to
the Pauli exclusion principle. For $\mathbf{k}_{1}\neq \mathbf{k}_{2}$, the
equation fulfilled by $G(\mathbf{k}_{1},\mathbf{k}_{2})$\ follows from the
cancellation of the above bracket. With probably in mind a $(\mathbf{\ k}%
_{1},\mathbf{k}_{2})$ decoupling , Richardson suggested to split $\mathcal{E}%
_{2}$\ as 
\begin{equation}
\mathcal{E}_{2}=R_{1}+R_{2}  \label{11}
\end{equation}%
with $R_{1}\neq R_{2}$, a requirement mathematically crucial as seen below.
We can then note that 
\begin{equation}
\frac{2\varepsilon _{\mathbf{k}_{1}}+2\varepsilon _{\mathbf{k}_{2}}-\mathcal{%
E}_{2}}{\left( 2\varepsilon _{\mathbf{k}_{1}}-R_{1}\right) \left(
2\varepsilon _{\mathbf{k}_{2}}-R_{2}\right) }=\frac{1}{2\varepsilon _{%
\mathbf{k}_{1}}-R_{1}}+\frac{1}{2\varepsilon _{\mathbf{k}_{2}}-R_{2}}
\label{12}
\end{equation}%
with ($R_{1},R_{2}$) possibly exchanged. This probably led Richardson to see
that the symmetrical function constructed on the LHS of the above equation,
namely%
\begin{equation}
G(\mathbf{k}_{1},\mathbf{k}_{2})=\frac{1}{\left( 2\varepsilon _{\mathbf{k}%
_{1}}-R_{1}\right) \left( 2\varepsilon _{\mathbf{k}_{2}}-R_{2}\right) }%
+(R_{1}\leftrightarrow R_{2})  \label{13}
\end{equation}%
is an exact solution of the Schr\"{o}dinger equation provided that $R_{1}$\
and $R_{2}$ are such that 
\begin{equation}
1=V\sum \frac{w_{\mathbf{k}}}{2\varepsilon _{\mathbf{k}}-R_{1}}+\frac{2V}{%
R_{1}-R_{2}}=(R_{1}\leftrightarrow R_{2})  \label{14}
\end{equation}%
as obtained by inserting Eq.(13) into $\left( H-\mathcal{E}_{2}\right)
\left\vert \psi _{2}\right\rangle =0$. Note that the denominator in the
above equation clearly shows why $(R_{1},R_{2})$ are required to be
different. The fundamental advantage of Richardson's procedure is to replace
the resolution of a 2-body Schr\"{o}dinger equation for $G(\mathbf{k}_{1},%
\mathbf{k}_{2})$\ by a problem far simpler, namely, the resolution of two
nonlinear algebraic equations.

This procedure nicely extends to $N$ pairs, the equations for $R_{1}$, ..., $%
R_{N}$ reading as Eq.(14), with all possible $R$ differences. However, to
the best of our knowledge, the analytical resolution of these equations for
arbitrary $N$ has stayed an open problem, even when $N=2$. We now show how
we can tackle this resolution analytically, first through a perturbative
approach, and then through two exact procedures.

\subsection{Perturbative approach}

A simple way to tackle the Richardson's equations analytically is to note
that Eq.(4) allows to replace 1 in the LHS of Eq.(14) by the same sum with $%
R_{1}$ replaced by $\mathcal{E}_{1}$. If we now add and substract the two
Richardson's equations, we get two equations in which the potential $V$ has
formally disappeared, namely%
\begin{equation}
\sum \left( \frac{w_{\mathbf{k}}}{2\varepsilon _{\mathbf{k}}-R_{1}}+\frac{w_{%
\mathbf{k}}}{2\varepsilon _{\mathbf{k}}-R_{2}}\right) =2\sum \frac{w_{%
\mathbf{k}}}{2\varepsilon _{\mathbf{k}}-\mathcal{E}_{1}}  \label{15}
\end{equation}%
\begin{equation}
\sum \left( \frac{w_{\mathbf{k}}}{2\varepsilon _{\mathbf{k}}-R_{1}}-\frac{w_{%
\mathbf{k}}}{2\varepsilon _{\mathbf{k}}-R_{2}}\right) =-\frac{4}{R_{1}-R_{2}}
\label{16}
\end{equation}%
$V$ is in fact hidden into $\mathcal{E}_{1}$. This is a wise way to put the
singular $V$ dependence of Cooper pairs into the problem, at minimum cost.

In view of Eq.(15), we are led to expand the sums appearing in Richardson's
equations as 
\begin{eqnarray}
\sum_{\mathbf{k}}\frac{w_{\mathbf{k}}}{2\varepsilon _{\mathbf{k}}-R_{1}}
&=&\sum_{\mathbf{k}}\frac{w_{\mathbf{k}}}{2\varepsilon _{\mathbf{k}}-%
\mathcal{E}_{1}+\mathcal{E}_{1}-R_{1}}  \notag \\
&=&\sum_{n=0}^{\infty }J_{n}(R_{1}-\mathcal{E}_{1})^{n}  \label{17}
\end{eqnarray}%
where $J_{0}=1/V$\ while $J_{n>0}$ is a positive constant given by%
\begin{equation}
J_{n}=\sum_{\mathbf{k}}\frac{w_{\mathbf{k}}}{\left( 2\varepsilon _{\mathbf{k}%
}-\mathcal{E}_{1}\right) ^{n+1}}=\frac{\rho _{0}}{2}\frac{I_{n}}{%
n\varepsilon _{c}^{n}}  \label{18}
\end{equation}%
\begin{equation}
I_{n}\simeq 1-e^{-2n/v}  \label{19}
\end{equation}%
for $v$ small. For this expansion to be valid, we must have $|R_{i}-\mathcal{%
E}_{1}|<2\varepsilon _{\mathbf{k}}-\mathcal{E}_{1}$ for all $k$. This
condition is going to be fulfilled for large samples, as possible to check
in the end.

It is convenient to look for $R_{i}$ through $C_{i}=\left( R_{i}-\mathcal{E}%
_{1}\right) /\varepsilon _{c}$\ with $i=(1,2)$. Eqs.(15, 16) then give 
\begin{equation}
\sum_{n=1}^{\infty }\frac{I_{n}}{n}(C_{1}^{n}+C_{2}^{n})=0  \label{20}
\end{equation}%
\begin{equation}
(C_{1}-C_{2})\sum_{n=1}^{\infty }\frac{I_{n}}{n}(C_{1}^{n}-C_{2}^{n})=-2%
\gamma _{c}  \label{21}
\end{equation}%
The above formulation evidences that the Richardson's equations contain a
small dimensionless parameter, namely 
\begin{equation}
\gamma _{c}=4/N_{c}  \label{22}
\end{equation}%
where $N_{c}=\rho _{0}\varepsilon _{c}$. Indeed, $N_{c}$ is just the pair
number from which pairs start to overlap. This makes $N_{c}$ large, and
consequently $\gamma _{c}$ small compared to 1, in the large sample limit.

For $\gamma _{c}=0$ , the solution of the above equations reduces to $%
C_{1}=C_{2}=0$, i.e., $\mathcal{E}_{2}=2\mathcal{E}_{1}$. The fact that the
two-pair energy $\mathcal{E}_{2}$ differs from the energy of two single
pairs $2\mathcal{E}_{1}$ is physically due to Pauli blocking, but
mathematically comes from a small but nonzero value of $\gamma _{c}$.

To solve Eqs.(20, 21) in the small $\gamma _{c}$ limit, it is convenient to
set $C_{1}=S+D$ and $C_{2}=S-D$. This allows us to rewrite Eqs.(20, 21) as 
\begin{eqnarray}
-D^{2} &=&  \label{23} \\
&&\frac{\gamma _{c}/2}{\left[ I_{1}\!+\!\frac{I_{3}}{2}(D^{2}\!+\!3S^{2})\!+%
\!\cdots \right] \!+\!S\left[ I_{2}\!+\!I_{4}(D^{2}\!+\!S^{2})\!+\!\cdots %
\right] }  \notag
\end{eqnarray}%
\begin{equation}
-S=\frac{\frac{I_{2}}{2}(D^{2}+S^{2})+\frac{I_{4}}{4}%
(D^{4}+6D^{2}S^{2}+S^{4})+\cdots }{I_{1}+\frac{I_{3}}{3}(3D^{2}+S^{2})+%
\cdots }  \label{24}
\end{equation}%
Their solution at lowest order in $\gamma _{c}$ reads $\gamma
_{c}/2I_{1}\simeq -D^{2}\simeq 2SI_{1}/I_{2}$. When inserted into $%
R_{1}+R_{2}=2\mathcal{E}_{1}+(C_{1}+C_{2})\varepsilon _{c}$, this gives the
two-pair energy as 
\begin{eqnarray}
\mathcal{E}_{2} &\simeq &2\mathcal{E}_{1}+\gamma _{c}\,\frac{I_{2}}{%
2I_{1}^{2}}\,\varepsilon _{c}  \notag \\
&\simeq &2\mathcal{E}_{1}+\frac{2}{\rho _{0}}\left( 1+2e^{-2/v}\right)
\label{25}
\end{eqnarray}%
Using the expression of $\mathcal{E}_{1}$ given in Eqs.(5,6), we can rewrite
this energy as 
\begin{eqnarray}
\mathcal{E}_{2} &\simeq &2\left[ \left( 2\varepsilon _{F_{0}}+\frac{1}{\rho
_{0}}\right) -\varepsilon _{c}\left( 1-\frac{1}{N_{\Omega }}\right) \right] 
\notag \\
&\simeq &2\left[ \left( 2\varepsilon _{F_{0}}+\frac{1}{\rho _{0}}\right)
-\varepsilon _{V}(N_{\Omega }-1)\right]  \label{26}
\end{eqnarray}%
Compared to the energy of two single pairs $2\mathcal{E}_{1}=2(2\varepsilon
_{F_{0}}-\varepsilon _{c})$, we see that Pauli blocking has two quite
different effects. (i) It first increases the normal part of this energy as
reasonable since the Fermi level for free electrons increases. The first
term in Eq.(26) is nothing but $2\varepsilon _{F_{0}}+2(\varepsilon _{F_{0}}+%
\frac{1}{\rho _{0}})$: one pair has a kinetic energy $2\varepsilon _{F_{0}}$
while the second pair has a slightly larger kinetic energy $2(\varepsilon
_{F_{0}}+\frac{1}{\rho _{0}})$, the Fermi level increase when one electron
is added, being $1/\rho _{0}$. (ii) Another less obvious effect of the Pauli
exclusion principle is to \textit{decrease} the average pair binding energy.
Indeed due to Pauli blocking, $(N_{\Omega }-1)$ pair states only are
available to form a bound state in the two-pair configuration, while all the 
$N_{\Omega }$ pair states are available in the case of a single Cooper pair.

\subsection{Exact approach}

The perturbative approach developed above, through the $\left( R_{i}-%
\mathcal{E}_{1}\right) $\ expansion of the sum appearing in the Richardson's
equations, helped us to easily get the effect of Pauli blocking on the
ground state of two Cooper pairs. It is in fact possible to avoid this $%
\gamma _{c}$ expansion as we now show.

Through the perturbative calculation, we have found that the difference $%
R_{1}-R_{2}$ is imaginary at first order in $\gamma _{c}$. It is possible to
prove that this difference is imaginary at any order in $\gamma _{c}$: The
Richardson's procedure amounts to add an imaginary part to the two-pair
energy, in order to escape into the complex plane and avoid poles in sums
like the one of Eq.(4), the two "Richardson's energies" then reading as $%
R_{1}=R+iR^{\prime }$ and $R_{2}=R-iR^{\prime }$, with $R$ and $R^{\prime }$
real.

$R$ is by construction real since $R_{1}+R_{2}=2R$ is the energy of the two
Cooper pairs. In order to show that $R^{\prime }$ also is real, let us go
back to Eq.(16). In terms of $(R,R^{\prime })$, this equation reads

\begin{equation}
\sum \frac{w_{\mathbf{k}}}{X_{\mathbf{k}}^{2}+R^{{\prime }2}}=\frac{1}{R^{{%
\prime }2}}  \label{27}
\end{equation}%
where $X_{\mathbf{k}}=2\varepsilon _{\mathbf{k}}-R$ is real. We then note
that this equation also reads

\begin{equation}
\sum \frac{w_{\mathbf{k}}X_{\mathbf{k}}^{2}}{\left\vert X_{\mathbf{k}%
}^{2}+R^{{\prime }2}\right\vert ^{2}}=R^{\prime }{}^{\ast 2}\left[ \frac{1}{%
\left\vert R^{\prime }\right\vert ^{4}}-\sum \frac{w_{\mathbf{k}}}{%
\left\vert X_{\mathbf{k}}^{2}+R^{{\prime }2}\right\vert ^{2}}\right]
\label{28}
\end{equation}%
From it, we readily see that, since the LHS and the bracket are both real, $%
R^{\prime }{}^{\ast 2}$ must be real. $R^{\prime }{}^{\ast 2}$ can then
either be positive or negative, e.i., $R^{\prime }$ can be real or
imaginary, which produces $(R_1,R_2)$ either both real or complex conjugate.

To show that $(R_{1},R_{2})$ cannot be both real, we go back to Eq.(16). By
noting that $\sum w_{\mathbf{k}}$ is nothing but the number $N_{\Omega }$ of
pairs in the potential layer, we can rewrite this equation as 
\begin{equation}
0=\sum w_{\mathbf{k}}\left[ \frac{1}{2\varepsilon _{\mathbf{k}}-R_{1}}-\frac{%
1}{2\varepsilon _{\mathbf{k}}-R_{2}}+\frac{4}{N_{\Omega }(R_{1}-R_{2})}%
\right]
\end{equation}%
\begin{equation}
=\frac{1}{R_{1}-R_{2}}\sum w_{\mathbf{k}}\frac{A_{\mathbf{k}}}{(2\varepsilon
_{\mathbf{k}}-R_{1})(2\varepsilon _{\mathbf{k}}-R_{2})}
\end{equation}%
where 
\begin{equation}
A_{\mathbf{k}}=(R_{1}-R_{2})^{2}+\frac{4}{N_{\Omega }}(2\varepsilon _{%
\mathbf{k}}-R_{1})(2\varepsilon _{\mathbf{k}}-R_{2})
\end{equation}%
It is possible to rewrite the second term of $A_{\mathbf{k}}$ using $%
4ab=(a+b)^{2}-(a-b)^{2}$. This leads to 
\begin{equation}
A_{\mathbf{k}}=(1-\frac{1}{N_{\Omega }})(R_{1}-R_{2})^{2}+\frac{1}{N_{\Omega
}}(4\varepsilon _{\mathbf{k}}-R_{1}-R_{2})^{2}
\end{equation}%
Since the number of pairs $N_{\Omega }$ in the potential layer is far larger
than $1$, $A_{\mathbf{k}}$ would be positive if $(R_{1},R_{2})$ were both
real. For $(R_{1},R_{2})$ outside the potential layer over which the sum
over $\mathbf{k}$ is taken, the sum in Eq.(30) would be made of terms with a
given sign, so that this sum cannot cancels. Consequently, solutions outside
the potential layer must be complex conjugate whatever $\gamma _{c}$. 


For $(R_1,R_2)$ complex conjugate, i.e., $R^{\prime }$ real, the sum over $%
\mathbf{k}$ in Eq.27, performed within a constant density of states, leads
to 
\begin{eqnarray}
\frac{1}{R^{\prime }{}^{2}} &=&\frac{\rho _{0}}{2}\int_{\varepsilon
_{F_{0}}}^{\varepsilon _{F_{0}}+\Omega }\frac{2d\varepsilon _{\mathbf{k}}}{%
X_{\mathbf{k}}^{2}+R^{\prime }{}^{2}}  \notag \\
&=&\frac{\rho _{0}}{2R^{\prime }}\left( \arctan \frac{2\Omega +X_{F_{0}}}{%
R^{\prime }}-\arctan \frac{X_{F_{0}}}{R^{\prime }}\right)  \label{30}
\end{eqnarray}%
where $X_{F_{0}}=2\varepsilon _{F_{0}}-R$. If we now take the tangent of the
above equation, we find

\begin{equation}
\tan \frac{2}{\rho _{0}R^{\prime }}=\frac{2{\Omega }R^{\prime }}{R^{\prime
}{}^{2}+X_{F_{0}}(2\Omega +X_{F_{0}})}  \label{31}
\end{equation}%
Turning to Eq.(15), we find that it reads in terms of $(R,R^{\prime })$ as

\begin{equation}
\sum \frac{X_{\mathbf{k}}}{X_{\mathbf{k}}^{2}+R^{\prime }{}^{2}}=\frac{1}{V}
\label{32}
\end{equation}%
If we again perform the integration over $\mathbf{k}$ with a constant
density of states, this equation gives

\begin{equation}
\frac{X_{F_{0}}^{2}+R^{\prime }{}^{2}}{(2\Omega +X_{F_{0}})^{2}+R^{\prime
}{}^{2}}=e^{-4/v}  \label{33}
\end{equation}%
$R$ and $R^{\prime }$ then appear as the solutions of two algebraic
equations, namely Eqs.(31) and (33). Unfortunately, they do not have compact
form solutions.

It is however possible to solve these equations analytically in the large
sample limit. $\rho _{0}$ then goes to infinity so that $N_{\Omega }$ and $%
N_{c}$ are both large. In this limit $\tan (2/\rho _{0}R^{\prime })\simeq
2/\rho _{0}R^{\prime }$ to lowest order in $(1/\rho _{0})$. Eq.(31) then
gives $R^{\prime }{}^{2}\simeq X_{F_{0}}(2\Omega +X_{F_{0}})/N_{\Omega }$.

For $\rho _{0}$ infinite, i.e., $N_{\Omega }$ infinite, $R^{\prime }$
reduces to zero so that, due to Eq.(33), $z=X_{F_{0}}/(2\Omega +X_{F_{0}})$
reduces to $e^{-2/v}$. Eq.(33) can then be rewritten as 
\begin{equation}
z^{2}\simeq e^{-4/v}\frac{1+z/N_{\Omega }}{1+1/zN_{\Omega }}  \label{34}
\end{equation}%
Since $e^{-2/v}N_{\Omega }=N_{c}/2$ is also large compared to 1, this gives
the first order correction in $1/\rho _{0}$ to $z\simeq e^{-2/v}$ as $%
z\simeq e^{-2/v}\left[ 1+(e^{-2/v}-e^{2/v})/2N_{\Omega }\right] $. From $%
X_{F_{0}}=2\Omega z/(1-z)$ which for $z$ small reduces to $X_{F_{0}}\simeq
2\Omega (z+z^{2})$, we end by dropping terms in $e^{-4/v}$, with $R$ at
first order in $1/\rho _{0}$ given by 
\begin{eqnarray}
R &\simeq &2\epsilon _{F_{0}}-2\Omega e^{-2/v}\left( 1-\frac{1}{N_{c}}-\frac{%
1}{N_{\Omega }}\right)  \notag \\
&\simeq &2\epsilon _{F_{0}}+\frac{1}{\rho _{0}}-\epsilon _{c}(1-\frac{1}{%
N_{\Omega }})  \label{35}
\end{eqnarray}%
Since $\mathcal{E}_{2}=2R$, this result is just the one obtained from the
perturbative approach given in Eq.(26).

The major advantage of this exact procedure is to clearly show that the
above result corresponds to the dominant term in both, the large sample
limit by dropping terms in $(1/\rho _{0})^{2}$ in front $1/\rho _{0}$, 
\textit{and} the small potential limit by dropping terms in $e^{-4/v}$ in
front of $e^{-2/v}$. As seen from the first expression of $R$ in Eq.(35),
the Pauli exclusion principle induces a double correction, in $1/N_{c}$ and
in $1/N_{\Omega }$ to the one-pair binding energy $\epsilon _{c}=2\Omega
e^{-2/v} $. However, the corrections in $1/N_{c}$ ends by giving a potential
free correction to the 2-pair energy $\mathcal{E}_{2}$ because, in a
non-obvious way, it in fact comes from a simple change in the free electron
Fermi sea filling, as seen from the second expression of $R$ in Eq.(35).

\subsection{Excited state}

We now consider the 2-pair excited states with a broken pair having a
nonzero total momentum, as possibly obtained by photon absorption. Such a
pair does not feel the BCS potential, so that it stays uncorrelated. These
excited states thus read%
\begin{equation}
\left\vert \psi _{1;\text{ }\mathbf{k,k}^{\prime }}\right\rangle =\sum F(%
\mathbf{k}_{1})\beta _{\mathbf{k}_{1}}^{\dagger }a_{\mathbf{k}\uparrow
}^{\dagger }a_{-\mathbf{k}^{\prime }\downarrow }^{\dagger }\left\vert
F_{0}\right\rangle  \label{36}
\end{equation}

To derive the equation fulfilled by $F(\mathbf{k}_{1})$, it is convenient to
note that, for $\mathbf{k\neq k}^{\prime }$, 
\begin{eqnarray}
&&\beta _{\mathbf{p}}\beta _{\mathbf{k}_{1}}^{\dagger }a_{\mathbf{k}\uparrow
}^{\dagger }a_{-\mathbf{k}^{\prime }\downarrow }^{\dagger }\left\vert
F_{0}\right\rangle  \notag \\
&=&\delta _{\mathbf{k}_{1}\mathbf{p}}\left( 1-\delta _{\mathbf{k}_{1}\mathbf{%
k}}-\delta _{\mathbf{k}_{1}\mathbf{k}^{\prime }}\right) a_{\mathbf{k}%
\uparrow }^{\dagger }a_{-\mathbf{k}^{\prime }\downarrow }^{\dagger
}\left\vert F_{0}\right\rangle  \label{37}
\end{eqnarray}%
the bracket insuring cancellation for $\mathbf{k}_{1}=\mathbf{k}$\ or $%
\mathbf{k}^{\prime }$, as necessary due to the LHS. It is then easy to show,
from the Schr\"{o}dinger equation $\left( H-\mathcal{E}_{1,\text{ }\mathbf{kk%
}^{\prime }}\right) \left\vert \psi _{1;\text{ }\mathbf{k,k}^{\prime
}}\right\rangle =0$ projected upon $\left\langle F_{0}\right\vert a_{-%
\mathbf{k}^{\prime }\downarrow }a_{\mathbf{k}\uparrow }\beta _{\mathbf{p}}$
that 
\begin{eqnarray}
0 &=&\left( 1-\delta _{\mathbf{pk}}-\delta _{\mathbf{pk}^{\prime }}\right) 
\notag \\
&&\left[ \left( 2\varepsilon _{\mathbf{p}}+\varepsilon _{\mathbf{k}%
}+\varepsilon _{\mathbf{k}^{\prime }}-\mathcal{E}_{1,\text{ }\mathbf{kk}%
^{\prime }}\right) F(\mathbf{p})\right.  \notag \\
&&\left. -Vw_{\mathbf{p}}\sum_{\mathbf{q\neq k,k}^{\prime }}w_{\mathbf{q}}F(%
\mathbf{q})\right]  \label{38}
\end{eqnarray}%
This makes $F(\mathbf{p})$ undefined for $\mathbf{p}=\mathbf{k}$ or $\mathbf{%
k}^{\prime }$. This is unimportant since the corresponding contribution in $%
\left\vert \psi _{1;\text{ }\mathbf{k,k}^{\prime }}\right\rangle $\ cancels
due to the Pauli exclusion principle. For $\mathbf{p}\neq (\mathbf{k}$, $%
\mathbf{k}^{\prime })$ the equation fulfilled by $F(\mathbf{p})$ is obtained
by enforcing the bracket of the above equation to cancel. Following the
one-Cooper pair procedure, we get the eigenvalue equation for one broken
pair $(\mathbf{k}$, $-\mathbf{k}^{\prime })$ plus one Cooper pair as%
\begin{equation}
\frac{1}{V}=\sum_{\mathbf{p\neq k,k}^{\prime }}\frac{w_{\mathbf{p}}}{%
2\varepsilon _{\mathbf{p}}+\varepsilon _{\mathbf{k}}+\varepsilon _{\mathbf{k}%
^{\prime }}-\mathcal{E}_{1,\text{ }\mathbf{kk}^{\prime }}}  \label{39}
\end{equation}

A first possibility is to have the two free electrons in the two lowest
states of the potential layer, namely $\varepsilon _{\mathbf{k}}=\varepsilon
_{F_{0}}$ and $\varepsilon _{\mathbf{k}^{\prime }}=\varepsilon
_{F_{0}}+1/\rho _{0}$. The $\mathbf{p}$-state energy in the above equation
must then be larger than $\varepsilon _{F_{0}}^{(2)}$ with $\varepsilon
_{F_{0}}^{(n)}=\varepsilon _{F_{0}}+n/\rho _{0}$; so that Eq.(39) merely
gives%
\begin{equation}
\frac{1}{V}=\frac{\rho _{0}}{2}\int_{\varepsilon
_{F_{0}}^{(2)}}^{\varepsilon _{F_{0}}+\Omega }\frac{2d\varepsilon }{%
2\varepsilon +2\varepsilon _{F_{0}}+1/\rho _{0}-\mathcal{E}_{1,\text{ }%
\mathbf{kk}^{\prime }}}  \label{40}
\end{equation}%
By writing $\varepsilon _{F_{0}}+\Omega $ as $\varepsilon
_{F_{0}}^{(n)}+\Omega ^{(n)}$ with $\Omega ^{(n)}=\Omega -n/\rho _{0}$,
Eqs.(4, 5) for the single pair energy readily give 
\begin{equation}
\mathcal{E}_{1,\text{ }\mathbf{kk}^{\prime }}-\left( 2\varepsilon _{F_{0}}+%
\frac{1}{\rho _{0}}\right) \simeq 2\left( \varepsilon _{F_{0}}+\frac{2}{\rho
_{0}}\right) -2\left( \Omega -\frac{2}{\rho _{0}}\right) e^{-2/v}  \label{41}
\end{equation}

Another possibility is to put the two free electrons in the second and third
lowest states of the potential layer, namely $\varepsilon _{\mathbf{k}%
}=\varepsilon _{F_{0}}+1/\rho _{0}$ and $\varepsilon _{\mathbf{k}^{\prime
}}=\varepsilon _{F_{0}}+2/\rho _{0}$. The $\mathbf{p}$-state energy in
Eq.(39) can then be equal to $\varepsilon _{F_{0}}$\ or larger than $%
\varepsilon _{F_{0}}^{(3)}$. In this case, Eq.(39) gives 
\begin{equation}
\frac{1}{V}=\frac{1}{2\varepsilon _{F_{0}}-E}+\frac{\rho _{0}}{2}%
\int_{\varepsilon _{F_{0}}^{(3)}}^{\varepsilon _{F_{0}}+\Omega }\frac{%
2d\varepsilon }{2\varepsilon -E}  \label{42}
\end{equation}%
in which we have set $E=\mathcal{E}_{1,\text{ }\mathbf{kk}^{\prime
}}-2\varepsilon _{F_{0}}-3/\rho _{0}$. By $E$ as $2\varepsilon
_{F_{0}}^{(3)}-2\Omega ^{(3)}e^{-2/v}x$, the above equation gives $x$ through%
\begin{equation}
\frac{2}{xN_{c}^{^{\prime \prime }}-3}=\mathrm{Log\,}\frac{x}{1+xe^{-2/v}}
\label{43}
\end{equation}%
where $N_{c}^{^{\prime \prime }}=2\Omega ^{(3)}e^{-2/v}\rho _{0}$ is close
to $N_{c}$, i.e., large compared to 1 in the large sample limit. This gives $%
x\simeq 1+2/N_{c}^{^{\prime \prime }}$, so that the energy $\mathcal{E}_{1,%
\text{ }\mathbf{kk}^{\prime }}$ would then be equal to%
\begin{equation}
\mathcal{E}_{1,\text{ }\mathbf{kk}^{\prime }}^{\prime }\simeq 4\varepsilon
_{F_{0}}+\frac{7}{\rho _{0}}-2\left( \Omega -\frac{3}{\rho _{0}}\right)
e^{-2/v}  \label{44}
\end{equation}%
This energy is larger than the one given in Eq.(41) with the broken pair in
the two lowest energy levels of the potential layer.

Such a conclusion stays valid for broken pair electrons in higher states:
the minimum energy for a broken pair plus a correlated pair is given by $%
\mathcal{E}_{1,\text{ }\mathbf{kk}^{\prime }}$ in Eq.(41). The excitation
gap to break one of the two Cooper pairs into two free electrons $\Delta =%
\mathcal{E}_{1,\text{ }\mathbf{kk}^{\prime }}-\mathcal{E}_{2}$, thus appears
to be 
\begin{equation}
\Delta =\varepsilon _{c}+\frac{3}{\rho _{0}}=\varepsilon _{c}\left( 1+\frac{3%
}{N_{c}}\right)  \label{45}
\end{equation}

We can then remember that the excitation gap for a single pair is equal to $%
\left[ \varepsilon _{F_{0}}+(\varepsilon _{F_{0}}+\frac{1}{\rho _{0}})\right]
-(2\varepsilon _{F_{0}}-\varepsilon _{c})$, i.e., $\varepsilon _{c}+\frac{1}{%
\rho _{0}}$: The broken pair being again in the \textit{two} lowest states
of the potential layer, this brings an additional $\frac{1}{\rho _{0}}$
contribution to the average pair binding energy $\varepsilon _{c}$. Eq.(45)
thus shows that the gap increases when going from one to two pairs. This
increase in fact comes from a mere kinetic energy increase induced by Pauli
blocking. It is worth noting that, while Pauli blocking induces an \textit{%
increase} of the gap, it produces a \textit{decrease} from $\varepsilon _{c}$
to $\varepsilon _{c}\left( 1-1/N_{\Omega }\right) $ of the average pair
binding energy when going from one to two correlated pairs. Since $N_{\Omega
}=\rho _{0}\Omega $\ is far larger than $N_{c}=\rho _{0}\varepsilon _{c}$,
the gap increase however is far larger than the binding energy decrease.

The changes we obtain in the excitation gap and in the average pair binding
energy when going from one to two pairs, are a strong indication that the
gap in the dense BCS configuration cannot be simply linked to the pair
binding energy, as commonly said. Indeed, the pair binding energy is going
to stay smaller than $\varepsilon _{c}=2\Omega e^{-2/v}$\ due to Pauli
blocking in the potential layer, while the experimental gap in the dense
regime is known to be of the order of $\Omega e^{-1/v}$ which is far larger
than $\varepsilon _{c}$.

We wish to stress that, in addition to the excited states considered in this
section, in which the broken pair ends by having a non-zero momentum, there
also are excited states, not included into the present work. In these
excited states, the two pairs still have a zero momentum but correspond to $%
R $'s located somewhere in the quasi-continuum spectrum of the one-electron
states, i.e., in-between two one-electron levels. For such $R$'s, it is not
possible to straightforwardly replace summation by integration in the
Richardson's equations as we did throughout the present paper.

\section{Conclusion}

We here extend the well-known one-pair problem, solved by Cooper, and
consider two correlated pairs of electrons added to a Fermi sea of
noninteracting electrons. The Schr\"{o}dinger equation for the two-pair
ground state has been reduced by Richardson to a set of two coupled
algebraic equations. We here give three different methods to solve these two
equations analytically in the large sample limit, providing a unique result.
These methods are perspective for the extension to an arbitrary number of
pairs in order to hopefully cover the crossover between dilute and dense
regimes of Cooper pairs, as well as to apply them to nanoscopic
superconductors. Although the two-pair problem we here solve, is only a
first step toward the resolution of this quite fundamental problem, it
already allows us to understand more deeply the role of Pauli blocking
between electrons from which pairs are constructed. We show that this
blocking leads to a \textit{decrease} of the average pair binding energy in
the two-pair system compared to the one-pair configuration. This decrease is
due to the fact that by increasing the number of pairs, we decrease the
number of available states to form bound pairs.

This two-pair problem actually has some direct relation to real physical
systems, where correlated pairs are more local than in conventional BCS
superconductors. We can mention underdoped cuprates, heavy fermion
superconductors, pnictides, and ultracold atomic gases. It was shown long
time ago\cite{Eagles}, in the context of superconducting semiconductors with
low carrier density, that the excitation spectrum of such dilute system of
pairs is controlled by the binding energy of an isolated pair rather than by
a cooperative BCS gap. Hence, this picture is very similar to the classical
Cooper model. Within the two-pair configuration that we here solve using
Richardson's procedure, we reach the same conclusion for the excitation
spectrum. We also reveal how the composite nature of correlated pairs
affects their binding energies through the Pauli exclusion principles for
elementary fermions from which the pairs are constructed.

The extrapolation of the tendency we find to the dense BCS regime of pairs,
indicates that the average pair binding energy in this regime must be
smaller than that of an isolated pair. At the same time, it is generally
believed that the pair binding energy in the BCS configuration is of the
order of the superconducting gap, which is much larger than the isolated
pair binding energy. To understand this discrepancy, we must note that there
are two rather different concepts of "pairs" in the many-particle BCS
configuration. Those with energy of the order of the gap are introduced not
ab initio, but enforced to have a gap energy in order to provide a
qualitative understanding for the expression for the ground state energy,
found within the BCS theory. These entities, called "virtual pairs" by
Schrieffer\cite{Schrieffer}, correspond to pairs of electrons \textit{excited%
} above the Fermi sea of noninteracting electrons. These virtual pairs have
to be contrasted with what Schrieffer calls "superfluid pairs"\cite%
{Schrieffer}, made of \textit{all} the electrons with opposite momenta
feeling the attracting BCS potential, the number of these pairs being much
larger than the number of "virtual pairs". Staying within the framework of
"superfluid pairs" greatly helps to understand the dilute and dense regimes
of pairs on the same footing.

\section{Acknowledgements}

W. V. P. acknowledges supports from the French Ministry of Education, RFBR
(project no. 09-02-00248), and the Dynasty Foundation.

\section{Appendix}

In this Appendix, we propose another exact approach to Richardson's
equations which may turn more convenient for problems dealing with a pair
number larger than two.

We start with Eqs.(14) and calculate the sum by again assuming a constant
density%
\begin{equation}
\frac{2}{v}=\int_{\varepsilon _{F_{0}}}^{\varepsilon _{F_{0}}+\Omega }\frac{%
2\,d\varepsilon }{2\varepsilon -R_{1}}+\frac{4}{\rho _{0}(R_{1}-R_{2})} 
\tag{A1}
\end{equation}%
with $R_{1}$ possibly complex. Instead of $R_{i}$, we are going to look for $%
a_{i}=(2\Omega +2\varepsilon _{F_{0}}-R_{i})/(2\varepsilon _{F_{0}}-R_{i})$\
with $i=(1,2)$. Since $R_{i}=2\varepsilon _{F_{0}}-2\Omega /(a_{i}-1)$, the
above equation yields 
\begin{equation}
\frac{2}{v}=\mathrm{Log\,}a_{1}+\frac{2}{N_{\Omega }}\frac{(a_{1}-1)(a_{2}-1)%
}{a_{1}-a_{2}}  \tag{A2}
\end{equation}%
where Log denotes the principal value of the complex logarithmic function,
i.e., the one that satisfies $-\pi <\mathrm{Im\,}(\mathrm{Log\,}a_{1})\leq
\pi $

By adding the same equation with $(1,2)$ exchanged, we readily get%
\begin{equation}
a_{1}a_{2}=e^{4/v}  \tag{A3}
\end{equation}%
This equation is nothing but Eq.(33), since for $R_{1}=R+iR^{\prime }$, we
do have $a_{1}=(2\Omega +X_{F_{0}}-R-iR^{\prime })/(X_{F_{0}}-R-iR^{\prime
}) $, with $i$ changed into $-i$ for ($R_{1}$, $a_{1}$) changed into ($R_{2}$%
, $a_{2}$). From Eq.(A3), we conclude that $a_{1}=e^{2/v}t$ while $%
a_{2}=e^{2/v}/t$.

Next, we note that, since the two-pair energy $\mathcal{E}_{2}=R_{1}+R_{2}$,
which also reads 
\begin{equation}
\mathcal{E}_{2}=4\varepsilon _{F_{0}}-2\Omega \frac{a_{1}+a_{2}-2}{%
e^{4/v}+1-(a_{1}+a_{2})}  \tag{A4}
\end{equation}%
is real, $(a_{1}+a_{2})$ must be real. This implies $t+1/t=t^{\ast
}+1/t^{\ast }$\ or equivalently $(t-t^{\ast })(tt^{\ast }-1)=0$.
Consequently, $t$ is either real or such that $\left\vert t\right\vert =1$,
i.e., $t=e^{i\varphi }$. To choose between these two possibilities, we
consider the difference of the two Richardson's equations as written in
Eq.(A2). This difference which first appears as 
\begin{equation}
0=\mathrm{Log\,}\frac{a_{1}}{a_{2}}+\frac{4}{N_{\Omega }}\frac{%
(a_{1}-1)(a_{2}-1)}{a_{1}-a_{2}}  \tag{A5}
\end{equation}%
reads in terms of $t$ as 
\begin{equation}
e^{2/v}+e^{-2/v}=t+\frac{1}{t}-\frac{N_{\Omega }}{2}\left( t-\frac{1}{t}%
\right) \mathrm{Log\,}t  \tag{A6}
\end{equation}%
For $t$ real, $(t-t^{-1})\mathrm{Log\,}t$ is always positive, except for $%
t=1 $ where it cancels. This shows that the RHS of Eq.(A6), equal to 2 for $%
t=1$, stays essentially smaller than 2 for $N_{\Omega }$ far larger than 1.
Since $e^{2/v}+e^{-2/v}$ is far larger than 2 for $v$ small, we conclude
that Eq.(A6) cannot be fulfilled for $t$ real.

The other possibility is $t=e^{i\varphi }$ with $0<|\varphi |<\pi $, so that
Log\thinspace $t=i\varphi $ We then have $a_{1}=a_{2}^{\ast }$, i.e., $%
R_{1}=R_{2}^{\ast }$. This shows that the two Richardson's energies are
complex conjugate, as found by the other exact approach. When used into
Eq.(A6), this $t$ leads to%
\begin{equation}
\varphi \sin \varphi +\tfrac{2}{N_{\Omega }}\cos \varphi =\delta _{c} 
\tag{A7}
\end{equation}%
with $\delta _{c}=\frac{e^{2/v}+e^{-2/v}}{N_{\Omega }}\simeq \frac{2}{N_{c}}$%
. Once Eq.(A7) for $\varphi $\ is solved, the two-pair energy given in
Eq.(A4) follows from%
\begin{equation}
\mathcal{E}_{2}=4\varepsilon _{F_{0}}-4\Omega \frac{e^{2/v}\cos \varphi -1}{%
e^{4/v}+1-2e^{2/v}\cos \varphi }  \tag{A8}
\end{equation}

The solutions of Eq.(A7) cannot be expressed in a compact form in terms of
classical functions. We however see that the two dimensionless terms in
Eq.(A7), namely $2/N_{\Omega }$ and $\delta _{c}$, are small. Furthermore $%
2/N_{\Omega }$ is smaller than $\delta _{c}$. The function in the LHS of
Eq.(A7) is increasing from $2/N_{\Omega }$ up to a maximum $\simeq 1.82$,
then decreasing down to $-2/N_{\Omega }$ as $\varphi $ runs in $[0,\pi ]$,
and still decreasing on $[\pi ,2\pi ]$. This shows that Eq.(A7) admits
exactly one solution in the interval $(0,\pi /2)$, another one in the
interval $(\pi /2,\pi )$ but no solution in $[\pi ,2\pi ]$. Changing $%
\varphi $ in $-\varphi $ would provide also two solutions on $(-\pi ,0)$ but
this just corresponds to exchange $a_{1}$ and $a_{2}$; so that they cannot
be considered as distinct solutions. For these solutions, $\varphi \sin
\varphi $\ stays close to zero, so that $\varphi $ is close to $0$ or to $%
\pi $. For $\varphi =0$, the RHS of the above equation reduces, for $v$
small, to $4\varepsilon _{F_{0}}-4\Omega e^{-2/v}$ which is just twice the
energy of a single Cooper pair as given in Eq. (6). The effect of Pauli
blocking on this two-single pair energy results from a large but finite
number of pair states $N_{\Omega }$ in the potential layer, as physically
expected. We can note that, by contrast, $\varphi =\pi $ would lead to $%
\mathcal{E}_{2}$ close to $4\varepsilon _{F_{0}}+4\Omega e^{-2/v}$. This
solution has to be sorted out because it corresponds to $R_{1}$ and $R_{2}$
located in the complex plane very close to the real axis where the
one-electron levels are positioned, so that the distance between them and
this real axis is of the order of $1/\rho _{0}$. This prevents substitution
of discrete summation by integration, in Eq.(A1), as discussed above.

For $\varphi $ close to zero, Eq.(A7) gives the leading term in $1/N_{\Omega
}$ as $\varphi ^{2}\simeq \left( e^{1/v}-e^{-1/v})^{2}\right/ N_{\Omega }$.
The ratio in Eq.(A8) then reads for $\varphi $ small as%
\begin{equation}
\frac{1}{e^{2/v}-1}\left[ 1-\frac{\varphi ^{2}}{2}\frac{e^{2/v}(e^{2/v}+1)}{%
(e^{2/v}-1)^{2}}\right]  \notag
\end{equation}%
\begin{equation}
\simeq \frac{1}{e^{2/v}-1}-\frac{e^{2/v}}{2N_{\Omega }}\frac{%
e^{4/v}-e^{2/v}-1+e^{-2/v}}{\left( e^{2/v}-1\right) ^{3}}  \tag{A9}
\end{equation}%
When inserted into Eq.(A8), we end with%
\begin{equation}
\mathcal{E}_{2}\simeq 4\varepsilon _{F_{0}}-4\Omega e^{-2/v}+\frac{2\Omega }{%
N_{\Omega }}\left( 1+2e^{-2/v}\right)  \tag{A10}
\end{equation}%
which is nothing but Eq.(26).

The main advantage of this second exact method is to have the two-pair
energy reading in terms of $\varphi $ which follows from a single equation,
namely Eq.(A7). By contrast, to get $\mathcal{E}_{2}$ through $%
X_{F_{0}}=2\varepsilon _{F_{0}}-\mathcal{E}_{2}$, as in the other exact
method, we must solve two coupled equations, namely Eqs.(31) and (33).

\end{document}